\UseRawInputEncoding
\documentclass{article}
\usepackage{appendix}
\usepackage[bookmarks=false]{hyperref}
\usepackage{amsmath}
\usepackage{amssymb}
\usepackage[retainorgcmds]{IEEEtrantools}
\usepackage{graphicx}

\begin{document}
\title{Condensation phenomena of ions in an electrostatic logarithmic trap}
\author{Loris Ferrari \\ Department of Physics and Astronomy (DIFA) of the University of Bologna\\via Irnerio, 46 - 40126, Bologna,Italy}
\maketitle
\begin{abstract}
The effects of an electrostatic logarithmic trap (ELT) on an ionic gas confined in a cylindric chamber are studied in detail, with special reference to the effects of the ion-ion Coulombic interactions and the resulting low-temperature thermodynamics. The collapse of the ions in radially localized states, about the axial cathode, is shown to cause an abrupt (but not critical) transition from non degeneration to strong degeneration, at a special temperature $T_c$. This transition could actually involve both Bosons and Fermions and is not to be confused with a Bose-Einstein condensation (BEC), which is excluded in principle. However, while for Bosons the resulting effects on the pressure are observable in the ultra high vacuum (UHV) regime, the Fermions' density should fall well below UHV, for the pressure change to be observable. This is because the ion-ion \emph{exchange} interactions increase the kinetic energy along the axial cathode, which makes the Fermi level and the non degeneration threshold temperature increase accordingly.         
\end{abstract}

\section{Introduction}
\label{intro}

Since 1963 \cite{Hoov}, the attention of theoreticians has been attracted by the dynamics of particles in a logarithmic potential \cite{Dechant,Guarnieri,Bouchaud,Aghion1,Aghion2,CEK}. In ref \cite{Me}, since now on referrred to as (I), the author studied a non interacting gas of $N$ particles, confined in a D-dimensional chamber of volume $V_D$ ($D=2,\:3$), under the action of an attractive potential $u(r)=u_0\ln(r/r_0)$ ($u_0>0$), denoted as \textquoteleft logarithmic trap\textquoteright$\:$(LT). It was shown therein that the critical temperature $T_B$ of Bose-Einstein condensation (BEC) in a LT diverges proportionally to $\ln V_D/\ln(\ln V_D)$ in the thermodynamic limit (TL), while a gas of Fermions, in the same conditions, exhibits a diverging Fermi level. In short: the ground state of LT's acts like a perfect attractor for any kind of particles at any temperature, in the TL. This is what the author called the \textquoteleft weirdness\textquoteright$\:$of a LT, with respect to other current cases of confinement by external fields, like the harmonic traps resulting from magneto-optical devices \cite{PS,Ph}. It is interesting to notice that the first hint to the peculiar behavior of LT's seems to date back to 1969, in an informal discussion with L. Onsager, reported in ref. \cite{Manning}.     

In principle, it is possible to arrange magnetic dipoles, or wind up coils in appropriate configurations, such as to realize a magnetic logarithmic trap (MLT). However, to the author's knowledge, no specific attempt has been made to test the feasibility of a MLT with appropriate size and strength. Other possible realizations of LT's for neutral particles are mostly hypothetical, like the attempt to deduce a hadronic mass formula \cite{Mur1,Mur2,Paa}, or refer to cosmological applications, like the Schwarzchild black holes \cite{Shak}.

In contrast to the magnetic option, an electrostatic logarithmic trap (ELT) is involved in the study of polyelectrolytes \cite{Manning} and in the standard technology widely applied to orbitrons and high-vacuum pumps \cite{Hoov,SD-M,Petit,Cyb}. In those cases, however, the ELT acts on \emph{charged} particles, and the non interacting approximation is definitely unsuitable, due to the long-range character of the Coulombic interparticle repulsion. The present work aims first to account for the inter-ionic repulsion, at the mean field level, in the study of \emph{firstly} ionized atoms, confined in the cylindric chamber sketched in Fig. 1. The axial wire of length $L$ and the cylindric wall of radius $R<<L$ are, respectively, the cathode and anode of an electric potential source, producing a 2D ELT, in the $\mathbf{x}$-plane normal to the $z$-axis (Section \ref{ELT}). It is assumed that the single-particle low-energy states of the ELT are strongly localized in the $\mathbf{x}$-plane, about the axial cathode, and free to move along the cathode itself. Following the Hatree-Fock (HF) method, it is shown that this \emph{ansatz} is self-consistent, since the resulting ion-ion repulsion contributes two separate terms to the single-particle potential at low temperatures: a logarithmic anti-trap, contrasting the ELT attraction, and a kinetic term $u_z(k)$, adding up to the free-particle kinetic energy $\hbar^2k^2/(2m)$ (Section \ref{HFB}). The former is a semiclassical consequence of the ions' localization about the axial cathode; the latter is a purely quantum effect resulting from the \emph{exchange} interactions.

\begin{figure}[htbp]
\begin{center}
\includegraphics[width=6in]{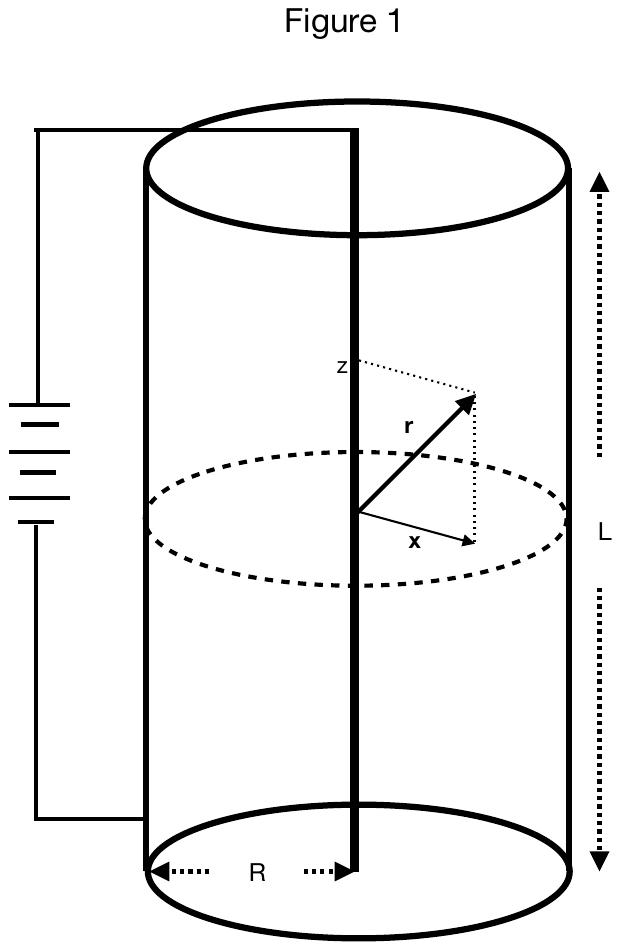}
\caption{\textbf{Schematic of the cylindric chamber hosting the ELT}. In the text, the radius $R$ is assumed small, compared to the length $L$, which is not reproduced in the figure for simplicity. The ion's position $\mathbf{r}=(\mathbf{x},\:z)$ is expressed in cylindric co-ordinates.}
\label{default}
\end{center}
\end{figure}

A further aim of the present work  is giving a detailed description of the ions thermodynamics in an ELT, an issue that is mostly ignored by the current literature on orbitrons and vacuum pumps, whose standard operative conditions actually make the effects of temperature's changes quite negligeable. As will be shown in what follows (Section \ref{PC}), the marked tendency of ELT's to push the gas in the lowest-energy states at any temperature (the \textquoteleft weirdness\textquoteright$\:$mentioned above), overshadows the effects of temperature changes, unless the particle density attains values typical of ultra-high vacua (UHV). In this case, despite BEC is excluded (Section \ref{noBEC}), the bosonic ions' gas exhibits a drastic change of the radial pressure at a special temperature $T_c$ (Section \ref{PC}), similar to - but not to be confused with - BEC. The pressure change is actually due to the abrupt - but not critical - passage from a non degenerate to a strongly degenerate regime, which is a further peculiar consequence of the logarithmic attraction, as discussed in (I). Actually, the same drastic change of the radial pressure might be observed, in principle, for ionic Fermions too, if it were not for the pure quantum term $u_z(k)$, resulting from the exchange interactions, which pushes the limit of concrete observability down to values of the Fermionic density well below the UHV range (Section \ref{PCFerm}).

\section{The Electrostatic Logarithmic Trap (ELT)}
\label{ELT}
In the chamber of Fig. 1, let $-|Q|/L$ and $|Q|/(2\pi LR)$ be the uniform charge densities on the axial wire and on the circular wall, respectively (the edge plates are assumed electrically neutral). In the vacuum technolgy, the ions - generated by the impact with fast electrons - experience the attractive force of the axial wire, in the $\mathbf{x}$-plane, and the corresponding repulsive force of the circular wall, both co-operating in pushing the ions towards the wire, where a ionic capture process finally occurs on the wire surface. Then each ion, assumed as a firstly ionized atom of charge $+e$, experiences an electrostatic potential expressed in cylindric coordinates $(\mathbf{x},\:z)$ as the sum of two terms $u_{wire}(x,\:z)$ and $u_{wall}(x,\:z)$ ($x:=|\mathbf{x}|$), due to the opposite charges distributed of the wire and on the wall:

\begin{subequations}
\label{uel0}
\begin{align}
u_{el}(x,\:z) &= \overbrace{u_0\int_{-L/2}^{L/2}\displaystyle\frac{\mathrm{d}z'}{\sqrt{(z'-z)^2+x^2}}}^{u_{wire}(x,\:z)}-\label{uwire}\\
\nonumber\\
&\underbrace{-u_0\int_{-L/2}^{L/2}\mathrm{d}z'\int_0^{2\pi}\displaystyle\frac{\mathrm{d}\theta}{\sqrt{(z'-z)^2+(R^2+x^2-2Rx\cos \theta)}}}_{u_{wall}(x,\:z)}\:,\label{uwall}
\end{align}
\end{subequations}
\\
where $u_0=2|Qe|/L$. From Appendix A, it follows that: 

\begin{subequations}
\begin{equation}
\label{uel1}
u_{el}(x,\:z) = \displaystyle u_0\ln\left(\frac{x}{L}\right)+u_0\ln(R/L)+\circ\left(R/L\right)+\circ\left(x/L\right)
\end{equation}
\\
for $x<R$, $|z|<L/2$, where $u_0=2|Qe|/L$. In the limit of divergingly large $L$, with $R<<L$, one can ignore the higher order terms. In addition, one can take advantage of the properties of the logarithm and write:

\begin{equation}
\label{uel2}
u_{el}(x,\:z) = u_0\ln\left(\frac{x}{x^*}\right)+u_0\ln(2x^*R^2/L^3)\quad(x<R<<L)\:,
\end{equation}
\end{subequations}
\\
where $x^*$ is an \emph{arbitrary} length scale, to be chosen at convenience. In particular, for log-trapped particles of mass $m$, an appropriate choice is $x^*=\hbar/\sqrt{2mu_0}$, as shown in (I).

\section{The mean field (MF) Hamiltonian from Hartree-Fock (HF) approach}
\label{HFB}

The interparticle coulombic interaction between two ions of charge $e$ reads:

\begin{equation}
\label{ucl}
u_{cl}(|\mathbf{r}-\mathbf{r}'|):=\frac{e^2}{\sqrt{(\mathbf{x}-\mathbf{x}')^2+(z-z')^2}}\:,
\end{equation}
\\
and yields the interparticle interaction energy:

\begin{equation}
\nonumber
U_{int}=\sum_{n=1}^{N}\sum_{j=n+1}^{N-1}u_{cl}(|\mathbf{r}_n-\mathbf{r}_j|)\:.
\end{equation}
\\
Let the states $\Psi_\xi(\mathbf{r}):=\langle\:\mathbf{r}\:|\:\xi\:\rangle$ form the (unknown) base for the construction of the $N$-bosons Fock states 

\begin{equation}
\nonumber
|\:\{N_\xi\:\}\:\rangle=\prod_\xi\frac{\left(b_\xi^\dagger\right)^{N_\xi}}{\sqrt{N_\xi!}}|\:\mathrm{vacuum}\:\rangle\:,
\end{equation}
\\
$N_\xi$ being the number of Bosons in the state $|\:\xi\:\rangle$, created by the operator $b_\xi^\dagger$. Following HF method, the mean interaction energy reads:

\label{UHF,dir,exc}
\begin{equation}
\nonumber
\langle\:\{N_\xi\:\}\:|U_{int}|\:\{N_\xi\:\}\:\rangle=\sum_{\xi}N_\xi\langle\:\xi\:|u_{\xi}|\:\xi\:\rangle\:,
\end{equation}
\\
where 

\begin{align}
\langle\:\xi\:|u_{\xi}|\:\xi\:\rangle&=\overbrace{\frac{1}{2}\sum_{\xi'}N_{\xi'}\int\mathrm{d}\mathbf{r}\int\mathrm{d}\mathbf{r}'u_{cl}(|\mathbf{r}-\mathbf{r}'|)|\Psi_{\xi'}(\mathbf{r}')|^2|\Psi_{\xi}(\mathbf{r})|^2}^{\text{direct interaction}}+\nonumber\\
\label{uxi}\\
&+\underbrace{\frac{1}{2}\sum_{\xi'}N_{\xi'}\int\mathrm{d}\mathbf{r}\:\Psi^*_{\xi}(\mathbf{r})\int\mathrm{d}\mathbf{r}'u_{cl}(|\mathbf{r}-\mathbf{r}'|)\Psi_{\xi'}(\mathbf{r})\Psi^*_{\xi'}(\mathbf{r}')\Psi_\xi(\mathbf{r}')}_{\text{exchange interaction}}\nonumber
\end{align}
\\
is the mean value of the interaction energy experienced by each ion in the $|\:\xi\:\rangle$ state. Equation \eqref{uxi} defines the operator $u_\xi$, whose action on the state $|\:\xi\:\rangle$ reads, in the co-ordinate representation: 

\begin{subequations}
\label{uxi2}
\begin{align}
\langle\:\mathbf{r}\:|u_{\xi}|\:\xi\:\rangle&=\frac{1}{2}\sum_{\xi'}N_{\xi'}\int\mathrm{d}\mathbf{r}'u_{cl}(|\mathbf{r}-\mathbf{r}'|)|\Psi_{\xi'}(\mathbf{r}')|^2\Psi_{\xi}(\mathbf{r})+\label{dir}\\
\nonumber\\
&+\frac{1}{2}\sum_{\xi'}N_{\xi'}(1-\delta_{\xi,\xi'})\Psi_{\xi'}(\mathbf{r})\int\mathrm{d}\mathbf{r}'u_{cl}(|\mathbf{r}-\mathbf{r}'|)\Psi^*_{\xi'}(\mathbf{r}')\Psi_\xi(\mathbf{r}')\:.\label{exc}
\end{align}
\end{subequations}   
\\
Due to the expression $\int\mathrm{d}\mathbf{r}'\cdots\Psi_\xi(\mathbf{r}')$, the exchange term \eqref{exc} turns out to be a \emph{non local} integral operator.

At low temperatures, the ions are expected to behave like free particles of wavevector $k$ along the $z$-axis, and to be strongly localized in the $\mathbf{x}$-plane, about the axial wire. The conditions for this assumption to apply will be found self-consistently in Section \ref{noBEC}. In particular, we set $\xi=(\eta,\:k)$, where $\eta=(n,\:m)$ labels the eigenstates of a radially symmetric Hamiltonian in 2D, and thereby includes the principal quantum number $n=0,\:1,\:2,\:\cdots$ and the orbital one $m=0,\:\pm1,\:\pm2,\:\cdots$. What precedes leads to the following \emph{ans\"atzen} about the $\Psi_\xi$'s:

\begin{itemize}
\item[(i)]
\begin{equation}
\nonumber
\Psi_\xi(\mathbf{x},\:z)=\phi_\eta(\mathbf{x})\frac{\mathrm{e}^{ikz}}{\sqrt{L}}\quad\text{for }|z|\le\frac{L}{2},\:\phi_\eta\text{ real}
\end{equation}
\item[(ii)]
\begin{equation}
\nonumber
\ell_\eta:=\int_{x<R}\mathrm{d}\mathbf{x}\:\phi_\eta^2(\mathbf{x})x<<R<<L\quad\text{for all the occupied }\eta\text{'s.}
\end{equation}
\\
\end{itemize}
The condition $\ell_\eta<<L$ yields a 1D free particle spectrum much denser than the low-energy spectrum $\{\epsilon_\eta\}$ in 2D. Actually, the separation between next neighbour low-energy levels scales like $\hbar^2/(mL^2)$, in the former case, and like $\hbar^2/(m\ell_0^2)$\footnote{$\eta=0$ is assumed as the ground state's label.} in the latter, so that the preliminary stages of degeneration are expected to exhibit a partial condensation of the ions in the lowest $\epsilon_\eta$-levels (localized in the $\mathbf{x}$-plane), leaving all the free particle levels still available for occupancy, along the $z$-axis. From \emph{ansatz} (i) and Eq. \eqref{ucl}, Equation \eqref{uxi2} becomes:

\begin{align}
&\langle\:\mathbf{x},\:z\:|u_{\eta,k}\:|\:\eta,\:k\:\rangle=\frac{e^2}{2L}\sum_{\eta',k'}N_{\eta',k'}\int_{\mathcal{B}}\frac{\mathrm{d}\mathbf{x}'\mathrm{d}z'}{\sqrt{(\mathbf{x}-\mathbf{x}')^2+(z-z')^2}}\times\nonumber\\
\label{umf}\\
&\times\left[\phi^2_{\eta'}(\mathbf{x'})\phi_{\eta}(\mathbf{x})+\phi_{\eta'}(\mathbf{x})\left(1-\delta_{k,k'}\delta_{\eta,\eta'}\right)\mathrm{e}^{i(k'-k)(z-z')}\phi_{\eta'}(\mathbf{x}')\phi_{\eta}(\mathbf{x}')\right]\:,\nonumber
\end{align}
\\
where $\mathcal{B}:=\left\{(\mathbf{x}',\:z');\:x'<R,\:|z'|<L/2\right\}$.\footnote{The supplemental condition $\sqrt{(\mathbf{x}-\mathbf{x}')^2+(z-z')^2}>d_m$, with $d_m\approx$ ionic diameter, ensures that the integral is convergent.} According to \emph{ansatz} (ii), it is possible to neglect $\mathbf{x}'$ in the square root in Eq. \eqref{umf}, \emph{modulo} terms of order $\ell_{min}/x$ (for $\phi_\eta(\mathbf{x})\phi_{\eta'}(\mathbf{x})$ odd), or $(\ell_{min}/x)^2$ (for $\phi_\eta(\mathbf{x})\phi_{\eta'}(\mathbf{x})$ even), $\ell_{min}$ being the minimum between $\ell_\eta$ and $\ell_{\eta'}$ (Appendix B). The condition $x>>\ell_{min}$ means that the inter-particle distances are large compared to the localization lenghts about the axial cathode. Therefore, neglecting $\mathbf{x}'$ in the square root of Eq. \eqref{umf} actually corresponds to a low-density approximation, according to which the integral resulting from the second (exchange) term in square brakets vanishes for $\eta\neq\eta'$, due to the orthogonality of $\phi_\eta$ and $\phi_{\eta'}$. Hence equation \eqref{umf} takes the form $\langle\:\mathbf{x},\:z\:|u_{\eta,k}\:|\:\eta,\:k\:\rangle=u_{\eta,k}(x,\:z)\Psi_{\eta,k}(\mathbf{x},\:z)$, with 

\begin{align}
u_{\eta,k}(x,\:z)&=\frac{e^2}{2L}\sum_{\eta',k'}N_{\eta',k'}\times\nonumber\\
\label{umf2}\\
&\times\left[\int_{-L/2}^{L/2}\frac{\mathrm{d}z'}{\sqrt{x^2+(z-z')^2}}+\left(1-\delta_{k,k'}\right)\int_{-L/2}^{L/2}\frac{\mathrm{d}z'\mathrm{e}^{i(k'-k)(z-z')}}{\sqrt{x^2+(z-z')^2}}\right]\:\nonumber
\end{align}
\\
a \emph{local} potential operator. In the limit $L\rightarrow\infty$, the first integral in Eq. \eqref{umf2} can be calculated according to the same method \cite{A} as in Section \ref{ELT} (recall Eq. \eqref{uwire}), while the second integral can be expressed \cite{Mth} in terms of a Bessel function as:

\begin{equation}
\nonumber
 2\times\mathrm{BesselK}[0,\:|k-k'|x]=2\left[\ln 2-\gamma_{Eu}-\ln(|k-k'|x)+\circ(x^2\ln(x|k-k'|)\right]\:,
 \end{equation}
 \\ 
where $\gamma_{Eu}=0.5772\cdots$ is Euler-Mascheroni constant. Neglecting the higher order terms in the preceding expression, and recalling that  $\sum_{\eta',k'}N_{\eta',k'}=N$, Equation \eqref{umf2} finally becomes:

\begin{align}
&u_{\eta,k}(x,\:z)=-\frac{2e^2N}{L}\ln\left(\frac{x}{x^*}\right)-\nonumber\\
\nonumber\\
&-\frac{e^2}{L}\sum_{\eta',k'}N_{\eta',k'}\left(1-\delta_{k,k'}\right)\left[\ln\left(\frac{x}{x^*}\right)+\ln(|k-k'|x^*)\right]+\text{ const}\:.\nonumber
\end{align}
\\
Notice that a term $\propto \ln(x^*/L)$ has been included in the constant, according to the same procedure as in Section \ref{ELT}. A rearrangement of the preceding equation yields:

\begin{subequations}
\label{umf,rho1,netak}
\begin{equation}
\label{umfa}
u_{\eta,k}(x,\:z)=u_x(x,\:k)+u_z(k)+\text{ const}\:,
\end{equation}
\\
where:

\begin{align}
&u_x(x,\:k):=-e^2\rho_1[3-\sum_{\eta'}n_{\eta',k}]\ln\left(\frac{x}{x^*}\right)\label{umfb}\\
\nonumber\\
&u_z(k):=-e^2\rho_1\sum_{\eta',k'}n_{\eta',k'}\left(1-\delta_{k,k'}\right)\ln(|k-k'|x^*)\:,\label{umfc}
\end{align}
\\

\begin{equation}
\label{rho1,netak}
\rho_1:=\frac{N}{L}\quad;\quad n_{\eta,k}:=\frac{N_{\eta,k}}{N}\:.
\end{equation}
\end{subequations}
\\

At the present level of approximation, the ion-ion interaction energy results in a logarithmic anti-trap potential $u_x(x,\:k)$, created by a \textquoteleft coat\textquoteright$\:$of ions wrapping the axial wire and contrasting the attractive ELT potential $u_{el}(x)$ (Eq. \eqref{uwire}). In addition to this effect, predictable even at a classical level, the ion-ion repulsion is shown to produce a positive energy term $u_z(k)$ (Eq. \eqref{umfb}), which adds up to the kinetic energy $\hbar^2k^2/(2m)$, resulting from the free motion along the $z$-axis. Interestingly, this kinetic term is totally due to the exchange interactions, and is, thereby, intrinsically quantum in nature.

\section{Self-consistency conditions in thermal equilibrium: absence of a genuine BEC}
\label{noBEC}

The approximations adopted so far have considerably simplified the problem, by removing the non linear character in the MF potential $u_\xi$. However, non linearity has not been dropped at all, but will come into play in the self-consistency conditions. Dropping the inessential constant, the single-particle Hamiltonian for the test ion in the state $\Psi_{\eta,k}(\mathbf{x},z)$ now reads:

\begin{equation}
\label{hetak}
H_{\eta,k}=\overbrace{-\frac{\hbar^2\bigtriangledown_{\mathbf{x}}}{2/m}+u_{el}(x)+u_x(x,\:k)}^{H_x}\underbrace{-\frac{\hbar^2\partial_z^2}{2m}+u_z(k)}_{H_z}\:.
\end{equation}
\\
Equation \eqref{hetak} is self-consistent with \emph{ansatz} (i), which requires that the potential energy experienced by the single ion is a constant in the axial co-ordinate $z$. Actually, the Schr\"odinger equation resulting from $H_z$ is authomatically solved by the free-particle expression $\mathrm{e}^{ikz}/\sqrt{L}$, with eigenvalues

\begin{equation}
\label{Tstr}
\mathcal{T}(k):=\frac{\hbar^2k^2}{2m}+u_z(k)-u_z(0)\:,
\end{equation}
\\
where the free particle lowest level $u_z(0)$ has been chosen as the energy origin. The separate Schr\"odinger equation $H_x\phi_\eta(\mathbf{x})=\epsilon_\eta\phi_\eta(\mathbf{x})$ determines the localized 2D states, and reads, from Eq.s \eqref{hetak}, \eqref{uel}, \eqref{umfa}:

\begin{equation}
\nonumber
\left[-\frac{\hbar^2\bigtriangledown_{\mathbf{x}}}{2m}+u_{eff}\ln(x/x^*)\right]\phi_\eta(\mathbf{x})=\epsilon_\eta\phi_\eta(\mathbf{x})\:,
\end{equation}
\\
where, recalling Eq. \eqref{umfb}:

\begin{equation}
\label{ueff}
u_{eff}:=u_0-\overbrace{e^2\rho_1[3-\sum_{\eta'}n_{\eta',k}]}^{u_{rep}}>0\:.
\end{equation}
\\
The total energy of the test ion is, of course:
 
\begin{equation}
\label{epsilonetak}
\epsilon_{\eta,k}=\epsilon_\eta+\mathcal{T}(k)\:.
\end{equation}
\\
The positivity condition in Eq. \eqref{ueff} is necessary, for the effective logarithmic potential to be attractive, and will be implicitly assumed since now on. Recalling Eq. \eqref{rho1,netak}, a further simplification comes from assuming that most of the ions populate the radial ground state $\phi_0(x)$, with eigenvalue $\epsilon_0$, which yields: 

\begin{subequations}
\begin{equation}
\label{ck}
n_{\eta,k}=n_k\delta_{\eta,0}\:,
\end{equation}
\\
\begin{equation}
\label{nk}
n_k=\frac{1}{N\left[\mathrm{e}^{\beta(\epsilon_0+\mathcal{T}(k)-\mu)}-1\right]}\:.
\end{equation}
\end{subequations}
\\
being the fraction of ions free to move along the axial wire, with wavevector $k$ and chemical potential $\mu$, in thermal equilibrium. Notice that, from Eq.s \eqref{nk} and \eqref{ueff}, the ground level $\epsilon_0$ would be itself a function of $k$, through $u_{rep}=e^2\rho_1(3-n_k)$ (Eq. \eqref{umfb}). However, the ions fraction $n_k$ is small to order $N^{-1}$, and vanishes in the thermodynamic limit ($N,\:L\rightarrow\infty$, $\rho_1<\infty$), unless a BEC temperature $T_B$ does exist, below which $n_0$ (the fractional population of the ground level) could achieve a non vanishing value. The calculations in what follows will show that BEC is excluded at any finite temperature self-consistently. Therefore, one can set
 
 \begin{equation}
 \label{urep}
 u_{rep}=3e^2\rho_1\:,
 \end{equation}
 \\
in the thermodynamic limit (TL), which removes the dependence on $k$ from the spectrum $\{\epsilon_\eta\}$ of the 2D localized states. Notice that 1/3 of the anti-trap effect is due to the \emph{exchange} ion-ion interaction.      

Apart from replacing $u_0$ with $u_{eff}=u_0-u_{rep}$, the radial Hamiltonian $H_x$ (Eq. \eqref{hetak}) is the same as the 2D Hamiltonian studied in (I), which makes it possible to use all the related results of (I) in the present work too. In particular, the ground state's radial part $\phi_0(x)$, the localization length $\ell_0$ and the ground level $\epsilon_0$ follow from a variational procedure, and read:

\begin{subequations}
\label{phix,alpha0,ell0}
\begin{equation}
\label{phix}
\phi_0(x)=\frac{\mathrm{e}^{-x/(2\ell_0)}}{\sqrt{\pi}}\quad,\quad\ell_0=\frac{\hbar}{\sqrt{m\:u_{eff}}}\quad,\quad\epsilon_0=u_{eff}\alpha_0\:,
\end{equation}
\\
\begin{equation}
\label{alpha0}
 \alpha_0=\frac{3-\ln 2}{2}-\gamma_{Eu}=0.576211\dots
\end{equation}
\end{subequations}
\\
($\gamma_{Eu}=$ Euler-Mascheroni constant). For the next developments, it is convenient to take the energy origin at the lowest energy value. Hence we redefine the Hamiltonian $H_x$ in Eq. \eqref{hetak} as

\begin{subequations}
\label{h,xc}
\begin{equation}
\label{hx}
h_x:=H_x-\epsilon_0=-\frac{\hbar^2\bigtriangledown_{\mathbf{x}}}{2m}+u_{eff}\ln\left(\frac{x}{x_c}\right)\:,
\end{equation}
\\
where
\begin{equation}
\label{xc}
x_c:=x^*\mathrm{e}^{\alpha_0}\sqrt{\frac{u_0}{u_{eff}}}=\ell_0\mathrm{e}^{\alpha_0}\sqrt{\frac{u_0}{2u_{eff}}}
\end{equation}
\end{subequations}
\\
is a rescaled localization length. Since $h_x$ has ground level zero, the whole energy spectrum of the ions producing the MF potential reduces to $\mathcal{T}(k)$. This yields, from Eq. \eqref{nk}:

\begin{equation}
\nonumber
n_k=\frac{1}{N\left[\mathrm{e}^{\beta(\mathcal{T}(k)-\mu)}-1\right]}\:,
\end{equation}
\\ 
which implies a non linear self-consistency equation for $\mathcal{T}(k)$, following from Eq.s \eqref{Tstr}, \eqref{umfc}:

\begin{subequations}
\label{Tstr2}
\begin{align}
\mathcal{T}(k)&=\frac{\hbar^2k^2}{2m}-\frac{e^2\rho_1}{N}\sum_{k'}\frac{(1-\delta_{k,k'})\ln(|k-k'|x^*)}{\left[\mathrm{e}^{\beta(\mathcal{T}(k')-\mu)}-1\right]}-u_k(0)=\nonumber\\
\nonumber\\
&=\frac{\hbar^2k^2}{2m}-\frac{e^2}{2\pi}\int_{-\infty}^{\infty}\mathrm{d}k'\frac{\left[\ln(|k-k'|x^*)-\ln(|k'|x^*)\right]}{\left[\mathrm{e}^{\beta(\mathcal{T}(k')-\mu)}-1\right]}=\label{Tstra}\\
\nonumber\\
&=\frac{\hbar^2k^2}{2m}-\frac{e^2}{2\pi}\int_0^\infty\mathrm{d}k'\ln (x^*k')\times\nonumber\\
\label{Tstrb}\\
&\times\left[\frac{1}{\left[\mathrm{e}^{\beta(\mathcal{T}(k-k')-\mu)}-1\right]}+\frac{1}{\left[\mathrm{e}^{\beta(\mathcal{T}(k'+k)-\mu)}-1\right]}-\frac{2}{\left[\mathrm{e}^{\beta(\mathcal{T}(k')-\mu)}-1\right]}\right]\nonumber
\end{align}
\end{subequations}
\\
where the continuum limit (CL) $\sum_k\cdots \rightarrow(L/(2\pi))\int\mathrm{d}k\cdots$ has been performed. Equations \eqref{Tstr2} show that $\mathcal{T}(k)=\mathcal{T}(-k)$ is an \emph{even} function of $k$. In the non degenerate limit $\beta\mu\rightarrow-\infty$, expression \eqref{Tstra} shows that $\mathcal{T}(k)\rightarrow\hbar^2k^2/(2m)$ recovers the pure free particle expression. Apart from corrections of order $u_k(0)/k^2$, this is the case for the limit $k\rightarrow\infty$ too, as shown by expression \eqref{Tstrb}. Finally, a series expansion in $k$ of expression \eqref{Tstrb} and the even parity of $\mathcal{T}(k)$ yield:

\begin{equation}
\label{Tstr0} 
\mathcal{T}(k)=k^2\left[\frac{\hbar^2}{2m}-\frac{e^2}{2\pi}\int_0^\infty\mathrm{d}k'\ln (x^*k')\frac{\mathrm{d}^2}{\mathrm{d}k'^2}\left(\frac{1}{\left[\mathrm{e}^{\beta(\mathcal{T}(k')-\mu)}-1\right]}\right)\right]+\circ(k^4)\:,
\end{equation}
\\
which shows that, for $\mu<0$ strictly, $\mathcal{T}(k)$ exhibit a free-particle like dependence $\propto k^2$ at small $k$ too. In this case, however, the proportionality coefficient corresponds to a temperature dependent effective mass, smaller than the nude ionic mass. Actually, the exchange term in Eq. \eqref{Tstr0} is expected to be positive, since it comes from the repulsive ion-ion interaction. No attempt will be made to solve Eq. \eqref{Tstr2} in the whole range of $k$-values. Equation  \eqref{Tstr0} is sufficient for the present aims, i.e. showing that a genuine BEC is excluded at any temperature. Indeed, for BEC to occur, the equation determining $\mu$:

\begin{equation}
\label{eqmu}
1=\sum_\eta\frac{1}{\pi\rho_1}\int_0^\infty\frac{\mathrm{d}k}{\left[\mathrm{e}^{\beta(\epsilon_\eta+\mathcal{T}(k)-\mu)}-1\right]}\:,
\end{equation}
\\
should have a finite solution $\beta_B=1/(\kappa T_B)$ for $\mu\rightarrow0$ and $\epsilon_\eta=0$. In this limit, however,  Equation \eqref{Tstr0} makes the integral in Eq. \eqref{eqmu} diverge at any finite $\beta$, which excludes BEC at all.

\section{A partial condensation in 2D}
\label{PC}

The results obtained so far are based on the consequences of \emph{ansatz} (ii), i.e. that the gap between the ground and the first excited level of the 2D spectrum $\{\epsilon_\eta\}$ is occupied by a quasi continuous spectrum of free-particle levels $\mathcal{T}(k)$. At sufficiently low temperatures, it is assumed that the gas undergo to a partial condensation of most of the ions in the 2D ground state $\phi_0(x)$ (Eq. \eqref{phix}), while the spectrum $\mathcal{T}(k)$ is thermally occupied by the free motion along the $z$-axis. As already mentioned, those moving ions form a sort of coat about the axial wire, which results in a logarithmic anti-trap (Eq. \eqref{umfb}), contrasting the ELT.  

The partial condensation in the lowest-energy states, localized in the $\mathbf{x}$-plane is \emph{not} a genuine phase transition, but an abrupt process, on the temperature scale, which simulates BEC fairly closely, as we shall see in what follows. We start from a high-temperature approximation, opposite to what has been explored so far, in which the continuum limit (CL) is adopted for the spectrum $\{\epsilon_\eta\}$ too, and the exchange contribution in $\mathcal{T}(k)$ is neglected:

\begin{equation}
\label{Tfree}
\mathcal{T}(k)\rightarrow\frac{\hbar^2k^2}{2m}\:,
\end{equation}
\\
according to the low degeneracy condition $\mathrm{e}^{-\beta\mu}>>1$. The density of states (DOS) corresponding to Eq. \eqref{Tfree},  is nothing but the free-particle DOS in 1D: 

\begin{equation}
\label{gz2}
g_z(\epsilon)=
\begin{cases}
\displaystyle\frac{L\sqrt{2m}}{\mathrm{h\sqrt{\epsilon}}}&\quad(\epsilon>0)\\
\\
0&\quad(\epsilon<0)
\end{cases}\:,
\end{equation}
\\
where $\mathrm{h}$ is the Planck constant. The DOS $g_x(\epsilon)$, resulting from the Hamiltonian \eqref{hx}, is exactly the same as the 2D case studied in (I), with suitable substitutions of symbols:

\begin{subequations}
\label{gx,S,epsc}
\begin{equation}
\label{gx2}
g_x(\epsilon)=\frac{\widehat{S}\mathrm{e}^{2\alpha_0}}{4u_{eff}}\times
\begin{cases}
0&\:\:\:\:(\epsilon\le0)\\
\\
\mathrm{e}^{2(\epsilon-\epsilon_c)/u_{eff}}&\:\:\:\:(0<\epsilon\le\epsilon_c)\\
\\
1&\:\:\:\:(\epsilon\ge\epsilon_c)
\end{cases}\:,
\end{equation}
\\
where $\widehat{S}$ is the base area measured in units of the microscopic area $\pi x_c^2$, and $\epsilon_c$ is the critical \emph{size-dependent} energy, diverging logarithmically with the base area $S=\pi R^2$:

\begin{equation}
\label{S,epsc}
\widehat{S}:=\left(\frac{R}{x_c}\right)^2\quad;\quad\epsilon_c:=\frac{u_{eff}}{2}\ln \widehat{S}\:.
\end{equation}
\end{subequations}
\\
The total DOS, $g(\epsilon)$ follows from the independent DOS's $g_x(\epsilon)$, $g_z(\epsilon)$, and from Eq.s \eqref{gx2}, \eqref{gz2}, with the aid of Appendix C:
 
\begin{subequations}
\label{g(eps)}
\begin{align}
&g(\epsilon)=\int_0^\epsilon\mathrm{d}\epsilon'g_z(\epsilon')g_x(\epsilon-\epsilon')=\nonumber\\ 
\nonumber\\
&=\mathcal{G}\times
\begin{cases}
J_-(\epsilon):=\displaystyle\sqrt{\frac{u_{eff}\pi}{8}}\mathrm{e}^{2\epsilon/u_{eff}}\mathrm{Erf}\left(\sqrt{2\epsilon/u_{eff}}\right)\quad&(0<\epsilon\le\epsilon_c)\\
\\
J_+(\epsilon):=\widehat{S}\sqrt{\epsilon-\epsilon_c}+\displaystyle\sqrt{\frac{u_{eff}\pi}{8}}\mathrm{e}^{2\epsilon/u_{eff}}\times\\
\\
\displaystyle\times\left[\mathrm{Erf}\left(\sqrt{2\epsilon/u_{eff}}\right)-\mathrm{Erf}\left(\sqrt{2(\epsilon-\epsilon_c)/u_{eff}}\right)\right]\quad&(\epsilon\ge\epsilon_c)\\
\end{cases}
\end{align}
\\
where

\begin{equation}
\label{G}
\mathcal{G}:=\frac{\mathrm{e}^{2\alpha_0}L}{\mathrm{h}\:u_{eff}}\sqrt{\frac{m}{2}}\:.
\end{equation}
\end{subequations}
\\
\begin{figure}[htbp]
\begin{center}
\includegraphics[width=5in]{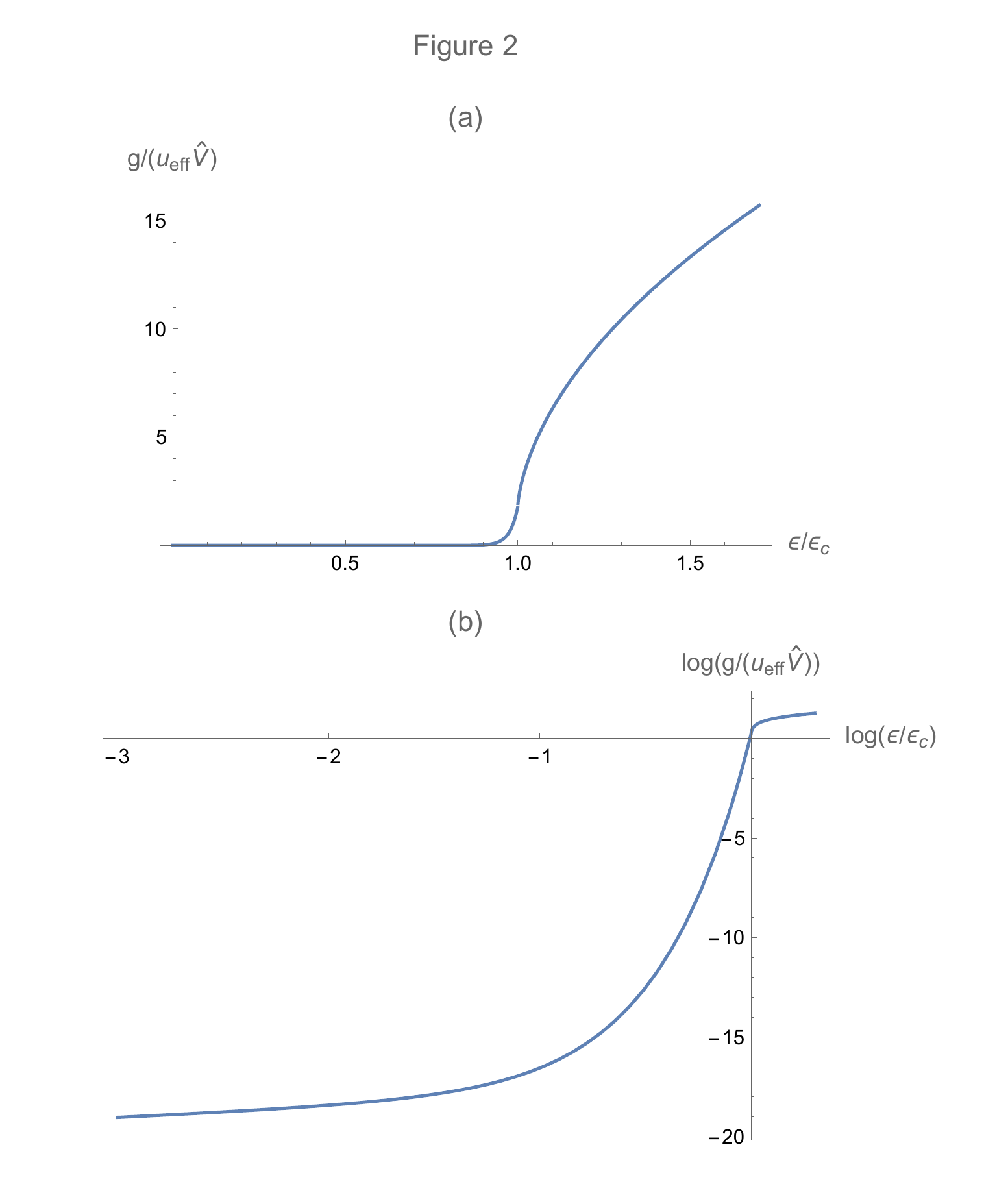}
\caption{\textbf{Semiclassical DOS of the ELT}. Linear plot (a) and log-log plot (b) of $g(\epsilon)$ as a function of $\epsilon/\epsilon_c$. $\widehat{V}:=SL/\ell_0^3$ is a dimensionless volume. The numerical parameters are $\ln S=43$, and $\mathcal{L}=10^3$ (Eq. \eqref{Lstr}).}
\label{default}
\end{center}
\end{figure}
The linear plot of $g(\epsilon)$ in Fig. 2a, shows the close analogy between the LT in a spherical 3D chamber, studied in (I), and the ELT in the cylindric chamber of Fig. 1. In this case too, the critical energy $\epsilon_c$ (Eq. \eqref{S,epsc}) acts like a gap from the lowest excited levels to the \emph{extensive³} free-particle DOS ($\propto \widehat{S}L\sqrt{\epsilon-\epsilon_c}$). However, the log-log plot in Fig. 3b shows that the DOS per unit volume  is not really zero for $\epsilon<\epsilon_c$, but small to order $1/V\propto1/(\widehat{S}L)$: $\epsilon_c$ would be a true gap only in the limit $V\rightarrow\infty$.\footnote{This is right the case studied in (I), where the log-trapped gas was considered in the thermodynamic limit $V,\:N\rightarrow\infty$, leading to the "weird" result of an infinite large gap $\epsilon_c\rightarrow\infty$, and a BEC at any temperature.}     

The equation:

 \begin{equation}
 \label{CP}
1=\frac{1}{N}\int_0^\infty\mathrm{d}\epsilon\frac{g(\epsilon)}{\mathrm{e}^{\beta(\epsilon-\mu)}-1}
\end{equation}
\\
yields the chemical potential $\mu$ for Bosons. From Eq. \eqref{g(eps)}, the series expansion of $1/(1-\mathrm{e}^{-\beta(\epsilon-\mu)})$ in equation \eqref{CP} yields:

\begin{subequations}
\begin{align}
\nonumber
1&=\frac{\mathcal{G}}{N}\left[\int_0^{\epsilon_c}\mathrm{d}\epsilon \frac{J_-(\epsilon)}{\mathrm{e}^{\beta(\epsilon-\mu)}-1}+\int_{\epsilon_c}^\infty\mathrm{d}\epsilon \frac{J_+(\epsilon)}{\mathrm{e}^{\beta(\epsilon-\mu)}-1}\right]=\nonumber\\
\label{mueq}\\
&=\mathcal{L}\sum_{n=1}^\infty\frac{\mathrm{e}^{n\beta\mu}}{(\theta n)^{3/2}}\left[\frac{1-\widehat{S}^{1-\theta n}}{(1-1/(\theta n))}+\frac{\pi}{8}\widehat{S}^{1-\theta n}\right]\:,\nonumber
\end{align}
\\
where
\begin{equation}
\label{Lstr}
\mathcal{L}:=\frac{\mathrm{e}^{2\alpha_0}\sqrt{\pi u_{eff}m}}{8\rho_1\mathrm{h}}\quad;\quad \theta:=\frac{T_c}{T}\quad;\quad T_c:=\frac{u_{eff}}{2\kappa}.
\end{equation}
\end{subequations}
\\
Equation \eqref{mueq} has been solved numerically, as reported in Fig. 3, showing a drastic change of the $\mu$'s slope at $T_c$. This follows from the first addendum ($n=1$) in the series of Eq. \eqref{mueq}, which is proportional to $\mathrm{e}^{\beta\mu+\ln \widehat{S}(1- T_c/T)}$, if $T\ge T_c$. Hence the limiting expression for the chemical potential reads:

\begin{equation}
\label{muB}
\mu\rightarrow-\kappa T\left(1-T_c/T\right)\ln\widehat{S}\quad\text{for }T> T_c\:;\:\ln\widehat{S}>>1\:.
\end{equation} 
\\
\begin{figure}[htbp]
\begin{center}
\includegraphics[width=4in]{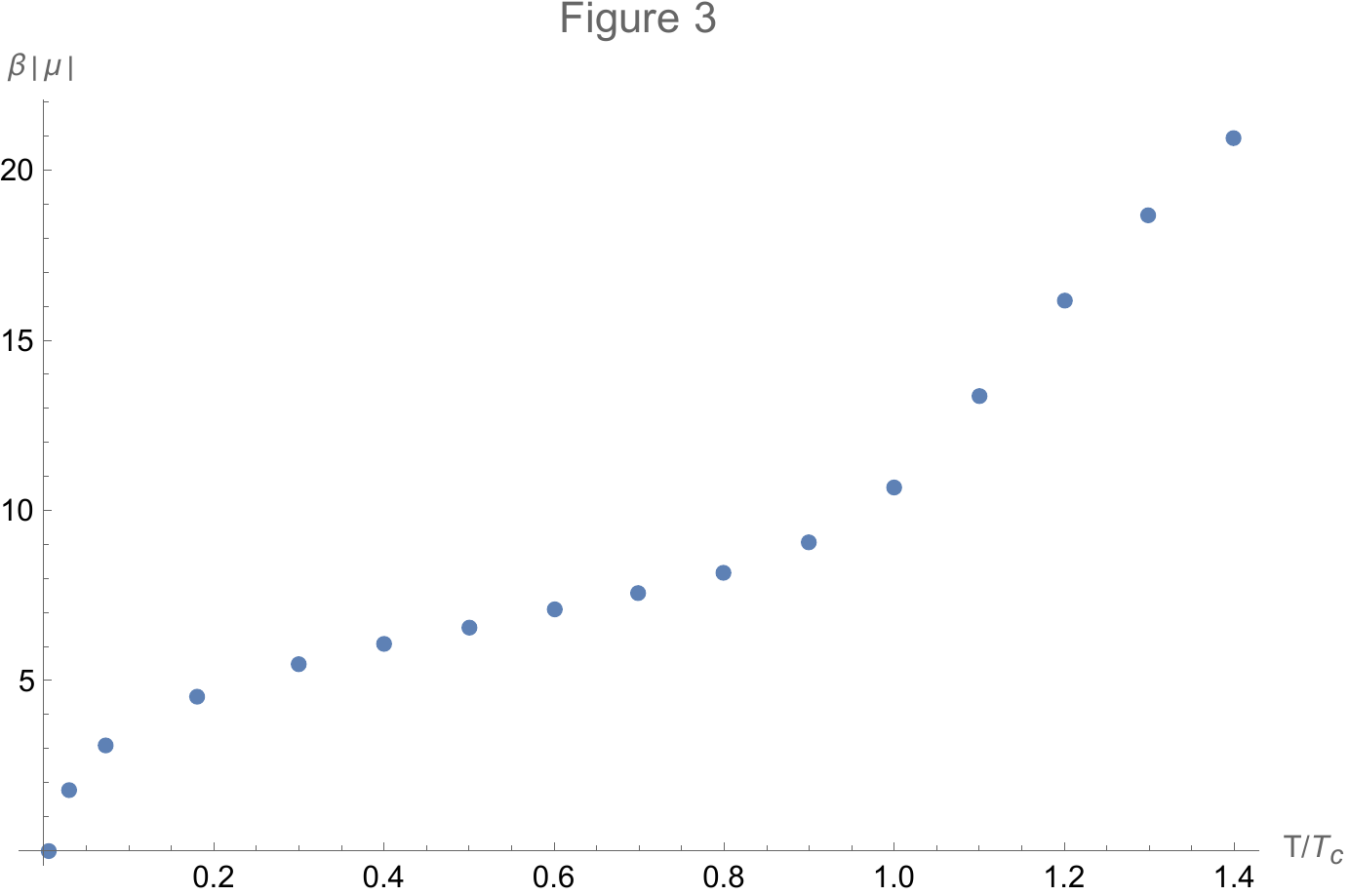}
\caption{\textbf{Bosonic chemical potential in the CL}. Plot of $|\mu|/(\kappa T)$ as a function of $T/T_c$. The numerical parameters are $\ln S=43$, and $\mathcal{L}=10^3$ (Eq. \eqref{Lstr}).}
\label{default}
\end{center}
\end{figure}

Equation \eqref{muB} shows that above the temperature $T_c:=u_{eff}/(2\kappa)$ the chemical potential becomes large negative, which means that the gas undergoes to a drastic transition from a highly degenerate to a non degenerate state, about $T_c$. This rapid escape from degeneration is due to the large factor $\ln \widehat{S}=\epsilon_c/(\kappa T_c)$, i.e. to the wide gap $\epsilon_c$ (Fig. 2a), characteristic of the logarithmic attraction. 

The cylindric symmetry and the non uniform spatial distribution of the ions yield two different expressions for the for the axial pressure, exerted by the gas on the chamber's sides: 

\begin{subequations}
\label{PBz,PBx}
\begin{equation}
\label{PBz}
P_z=-\frac{\kappa T}{S}\frac{\partial}{\partial L}\int_0^\infty\mathrm{d}\epsilon\:g(\epsilon)\log\left[1-\mathrm{e}^{-\beta(\epsilon-\mu)}\right]\:,
\end{equation}
\\
and for the radial pressure $P_x$, exerted on the circular wall:

\begin{equation}
\label{PBx}
P_x=-\frac{\kappa T}{S}\frac{\partial}{\partial S}\int_0^\infty\mathrm{d}\epsilon\:g(\epsilon)\log\left[1-\mathrm{e}^{-\beta(\epsilon-\mu)}\right]\:.\end{equation}
\end{subequations}
\\
Using Eq. \eqref{g(eps)} in Eq.s \eqref{PBz,PBx}, a numerical calculation, supported by the values of $\mu$ (Fig. 3), yields a semi-classical result $P_z\approx\rho_1\kappa T$, in agreement with the non degenerate limit adopted for $\mathcal{T}(k)$ (Eq. \eqref{Tfree}). However, when applied to Eq. \eqref{PBx}, the same calculation leads to Fig. 4, which shows a drastic change of $P_x$ at $T_c$. In particular, Figure 4(a) shows that $P_x(T)\approx \rho_3(T-T_c)$ for $T>T_c$, just like a semiclassical gas, but with the absolute zero shifted to $T_c$. Figure 4(b) shows the exponential fall of $P_x(T)$ for $T<T_c$. This illustrates very clearly the effects of the abrupt transition from a non degenerate to a strongly degenerate regime, which is what one expects in the ELT, when the effective volume occupied by the ions collapses from $SL$ to $L\pi \ell_0^2$, due to the partial condensation in the 2D ground state $\phi_0(x)$.

\begin{figure}[htbp]
\begin{center}
\includegraphics[width=3.5in]{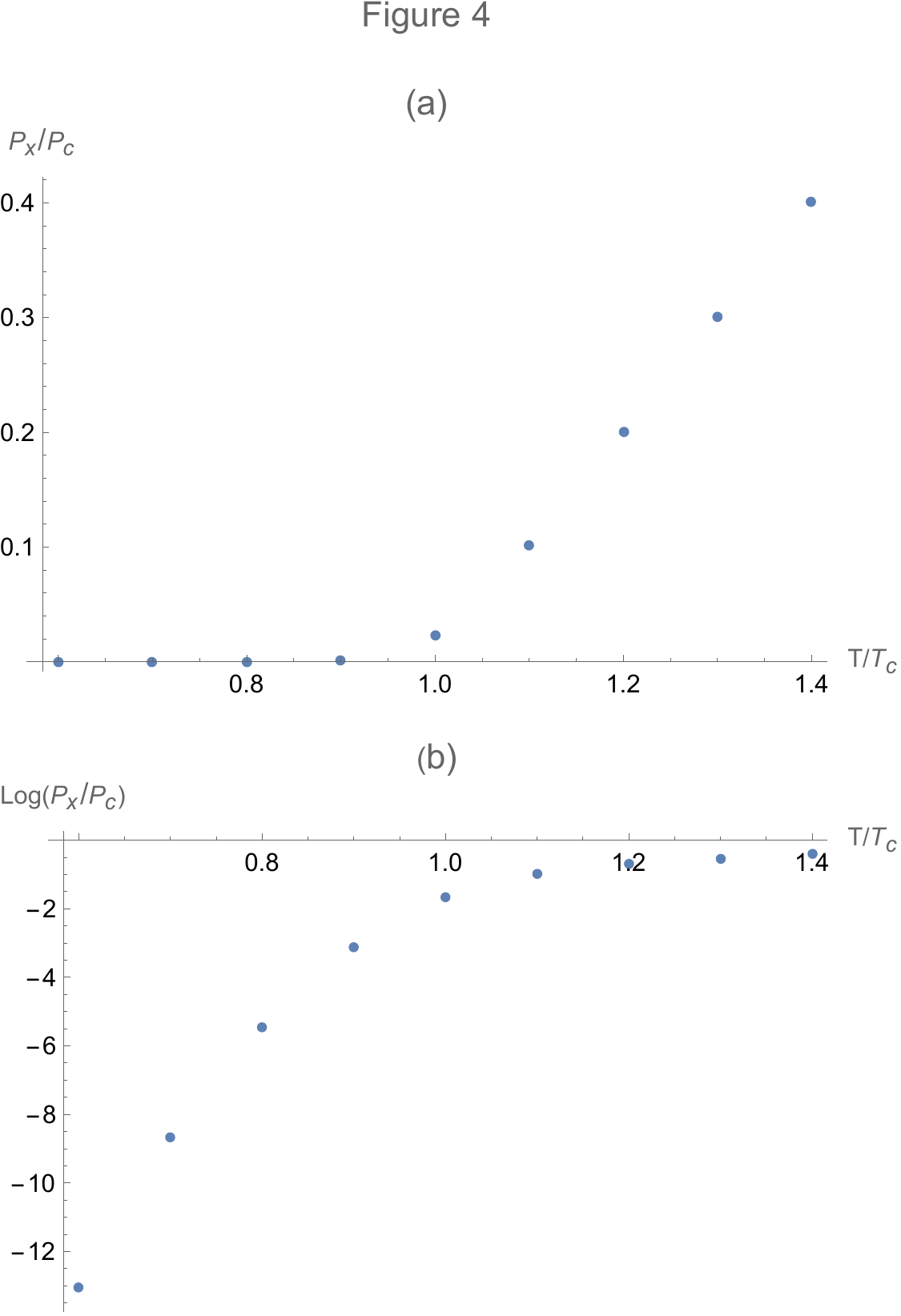}
\caption{\textbf{Bosonic radial pressure $P_x$ as a function of $T/T_c$}. Linear plot (a). Log-plot (b). $P_c:=\rho_3\kappa T_c$.}
\label{default}
\end{center}
\end{figure}    

The vanishing of $\mu$, shown in Fig. 3, simulates a false BEC at a finite temperature $T_{f}$, despite the results of Section \ref{noBEC} do exclude any genuine BEC, in an ELT. Actually, $T_f>0$ follows from a semiclassical approximation in which the discrete nature of the 2D localized spectrum $\{\epsilon_\eta\}$ is ignored. Therefore, $T_f$ marks a temperature scale below which the  gas is necessarily strongly degenerate (though not Bose-Einstein condensed). If $T_f\gtrsim T_c$, the drastic fall of the radial pressure shown in Fig. 4 does not occur, since the gas is  degenerate at any temperature, and the partial condensation described by (i) and (ii) occurs at any temperature too. If, instead: 

\begin{equation}
\label{SCC}
T_f<<T_c\:,
\end{equation}
\\ 
the drastic change of the radial pressure's slope is observable, due to the abrupt transition between the low and high degeneration regime. In particular, $T_f/T_c\approx5\times10^{-3}$, in Fig. 3. The equation for $T_f$ follows from Eq.s \eqref{CP}, \eqref{mueq} under the condition $\mu=0$ and $n_0=0$:

\begin{equation}
\label{thetaf}
1=\frac{1}{N}\int_0^\infty\mathrm{d}\epsilon\frac{g(\epsilon)}{\mathrm{e}^{\epsilon/\kappa T_f}-1}=\frac{\zeta(3/2)\mathcal{L}}{\theta_f^{3/2}}[1+\circ(1/\theta_f)]\:,
\end{equation}
\\
where $\theta_f:=2u_{eff}/(\kappa T_f)$ and $\zeta(\cdot)$ is Riemann's Zeta function. Notice that condition \eqref{SCC} is equivalent to $\theta_f>>1$. In this case, the dominant term in Eq. \eqref{thetaf} yields:

\begin{equation}
\label{Tf}
\kappa T_f\approx\frac{u_{eff}}{2}\left[\zeta(3/2)\mathcal{L}\right]^{-2/3}\quad;\quad\theta_f=[\zeta(3/2)\mathcal{L}]^{2/3}\:.
\end{equation}
\\

In what follows, we give a range of values for the physical parameters involved, in view of a concrete observation of the processes described above.  We use \textquoteleft tilded\textquoteright$\:$ dimensionless quantities: $\widetilde{Q}=Q/\text{unit}$. The ionic mass will be expressed in units of the Hydrogen mass, lenghts in centimeters, energies in $\mathrm{eV}$, temperatures in Kelvin. For the readers' benefit, the relevant expressions are collected in Table 1.
\\
\\
\begin{tabular}[t]{|c|c||c|c|}
\hline
\multicolumn{4}{|c|}{Table 1}\\
\hline
Formula & Meaning &  Formula & Meaning  \\
\hline\hline
&&&\\
\small{$S=\pi R^2$} & \footnotesize{Base Area} &\small{$u_{rep}=3e^2\rho_1$} &\footnotesize{Coulombic}  \\
&&&\footnotesize{rep. strength}\\
&&&\\
\hline
&&&\\
\small{$\rho_1=N/L$} & \footnotesize{Linear density} &\small{$\rho_3=N/(LS)$} & \footnotesize{Volume density}\\
&&&\\
\hline
&&&\\
\small{$u_0$} & \footnotesize{ELT} & \small{$u_{eff}=u_0-u_{rep}$} & \footnotesize{Effective} \\
&\footnotesize{strength} && \footnotesize{ELT strength}\\
&&&\\
\hline
&&&\\
\small{$\ell_0=\hbar/\sqrt{mu_{eff}}$}& \footnotesize{Radial} &\small{$x_c=\ell_0\mathrm{e}^{\alpha_0}\sqrt{u_0/(2u_{eff})}$}&\footnotesize{Rescaled} \\
&\footnotesize{loc. length}&&\footnotesize{loc. length}\\
&&&\\
\hline
&&&\\
\small{$\widehat{S}=(R/x_c)^2$} &\footnotesize{Dimensionless}  &\small{$\epsilon_c=u_{eff}\ln(\widehat{S})/2$} & \footnotesize{Effective} \\
&\footnotesize{base area}&&  \footnotesize{trap depth (gap)}\\
&&&\\
\hline
\end{tabular}
\\
\newline
\newline
From Table 1 and Eq.s \eqref{Lstr}, \eqref{Tf}, it is easy to get the following expressions :

\begin{subequations}
\begin{align}
\ln\widehat{S}&=41.08+\ln\left(\widetilde{R}^2\widetilde{m}\widetilde{u}_{eff}\right)\label{a}\\
\nonumber\\
\mathcal{L}&=1.732\times10^8\left(\frac{\widetilde{u}_{eff}\widetilde{m}}{\widetilde{\rho}_3^2\widetilde{R}^4}\right)^{1/2}\quad\quad\quad\:\label{b}\\
\nonumber\\
\widetilde{T}_f&=0.021\left(\frac{\widetilde{u}_{eff}\widetilde{\rho}_3\widetilde{R}^2}{m^{1/2}}\right)^{2/3}\quad\quad\quad\:\label{c}\:.
\end{align}
\end{subequations}
\\

First, we notice that $\ln\widehat{S}$ is large, as implicitly assumed all through the work, unless $u_{eff}$ and/or $R$ do assume unreasonably small values. In a typical high vacuum (HV) environment, one has $\widetilde{\rho}_3\approx10^9\div10^{13}$, and $\widetilde{u}_{eff}\approx10^3$. With a chamber radius $\widetilde{R}\approx1$ and a ionic mass $\widetilde{m}\approx10$, equation \eqref{c} yields enormously high values $\widetilde{T}_f\approx10^5\div10^7$, typical of plasmas. Hence, in standard HV conditions, one has $\theta_f<<1$, which means that the ionic gas of Bosons is totally collapsed in a \textquoteleft coat\textquoteright, wrapping the axial wire at any temperature. Actually, this is the goal the vacuum chambers aim to get, in ordinary operative conditions. In view of observing the collapse of the radial pressure $P_x(T)$ on the temperature scale, it is convenient to take $\widetilde{u}_{eff}\approx10^{-(1\div2)}$, which yields $T_c\lesssim\text{room temperature}$,  and deduce the resulting conditions on the other parameters, in particular the volume density, in order that the necessary condition $\theta_f>>1$ (Eq. \eqref{SCC}) is satisfied:

\begin{equation}
\label{rho1}
\widetilde{T}_c\approx10^2\div10^3\:;\:\theta_f>>1\quad\Rightarrow\quad\widetilde{\rho}_3<<\frac{\widetilde{m}}{\widetilde{R}^2}^{1/2}\times10^{8\div9}\:.
\end{equation} 
\\
With $\widetilde{R}\approx1$ and $\widetilde{m}\approx10$, one sees that the feasible range of densities is typical of ultra-high vacuum (UHV) environments ($\widetilde{\rho}_3\approx10^4\div10^9$).

\section{2D partial condensation of Fermions}
\label{PCFerm}

The study of a fermionic gas in a ELT follows the same line as in the bosonic case. In particular, we assume that at low temperatures each occupied eigenstate $|\:\xi\:\rangle$ of the MF single-particle Hamiltonian contains a pair of 1/2-spin Fermions. According to HF, the mean interaction energy experienced by the test ion reads:

\begin{subequations}
\label{uxif,nxif}
\begin{align}
\langle\:\xi\:|u_{\xi}|\:\xi\:\rangle&=\overbrace{\sum_{\xi'}N(\xi')\int\mathrm{d}\mathbf{r}\int\mathrm{d}\mathbf{r}'u_{cl}(|\mathbf{r}-\mathbf{r}'|)|\Psi_{\xi'}(\mathbf{r}')|^2|\Psi_{\xi}(\mathbf{r})|^2}^{\text{direct interaction}}-\nonumber\\
\label{uxif}\\
&\underbrace{-\frac{1}{2}\sum_{\xi'} N(\xi')\int\mathrm{d}\mathbf{r}\:\Psi^*_{\xi}(\mathbf{r})\int\mathrm{d}\mathbf{r}'u_{cl}(|\mathbf{r}-\mathbf{r}'|)\Psi_{\xi'}(\mathbf{r})\Psi^*_{\xi'}(\mathbf{r}')\Psi_\xi(\mathbf{r}')}_{\text{exchange interaction}}\:,\nonumber
\end{align}
\\
where

\begin{equation}
\label{nxif}
N(\xi)=\frac{2}{\left[\mathrm{e}^{\beta(\epsilon_\xi-\mu)}+1\right]}
\end{equation}
\end{subequations}
\\
is the occupation factor of the level $\epsilon_\xi$. Adopting the same \emph{ans\"atzen} and the same procedure as in Section \ref{HFB}, the factorization $\langle\:\mathbf{r}\:|\:\xi\:\rangle=\Psi_\xi(\mathbf{x},\:z)=\phi_\eta(\mathbf{x})\mathrm{e}^{ikz}/\sqrt{L}$ yields $\xi=(\eta,\:k)$, and the low-temperature approximation results in assuming that the large majority of Fermions occupy the 2D ground state $\phi_0(x)$, localized in the $\mathbf{x}$-plane, with localization length $\ell_0$ much smaller than $R$. This assumption makes it possible to save just the terms with $\xi'=(0,\:k')$, $\xi=(0,\:k)$ in Eq. \eqref{uxif}, which yields:

\begin{equation}
\langle\:\xi\:|u_{\xi}|\:\xi\:\rangle\rightarrow\langle\:k,\:0\:|u_{0,k}(x,\:z)|\:0,\:k\:\rangle\:.\nonumber
\end{equation}
\\
Following the same procedure as in Section \ref{noBEC}, the local operator expressing the MF single-particle potential finally reads, from Eq. \eqref{nxif}:

\begin{align}
u_{0,k}(x,\:z)&=\sum_{k'}\frac{e^2}{L}\int_{-L/2}^{L/2}\mathrm{d}z'\frac{N(0,\:k')}{\sqrt{x^2+(z'-z)^2}}\left[1-\frac{\mathrm{e}^{i(k-k')(z'-z)}}{2}\right]=\nonumber\\
\nonumber&\label{uxif2}\\
&=-3e^2\rho_1\ln(x/x^*)\underbrace{+\frac{e^2}{\pi}\int_{0}^\infty\mathrm{d}k'\frac{\ln(|k-k'|x^*)}{\mathrm{e}^{\beta(\mathcal{T}(k')-\mu)}+1}}_{u_z(k)}+\text{ constant}\:,
\end{align}
\\
where $\mathcal{T}(k)=\hbar^2k^2/(2m)+u_z(k)-u_z(0)$ is the 1D energy spectrum, with ground level $\mathcal{T}(0) =0$. What precedes yields the same ELT's effective strength $u_{eff}=u_0-u_{rep}$ (Eq. \eqref{urep}) as in the bosonic case. The equation for $\mathcal{T}(k)$ reads, according to Eq. \eqref{uxif2}:

\begin{equation}
\mathcal{T}(k)=\frac{\hbar^2k^2}{2m}+\frac{e^2}{\pi}\int_{-\infty}^{\infty}\mathrm{d}k'\frac{\left[\ln(|k-k'|x^*)-\ln(|k'|x^*)\right]}{\left[\mathrm{e}^{\beta(\mathcal{T}(k')-\mu)}+1\right]}\:.\label{Tstraf}
\end{equation}
\\
The Fermi level $\epsilon_F$ can be calculated from the limit $\beta\rightarrow\infty$ of Eq. \eqref{Tstraf}, which leads to the limiting expression of the 1D spectrum:

\begin{align}
\mathcal{T}_0(k)&:=\frac{\hbar^2k^2}{2m}+\frac{e^2}{\pi}\int_{|k'|<k_F}\mathrm{d}k'\left[\ln(|k-k'|x^*)-\ln(|k'|x^*)\right]=\nonumber\\
\nonumber\\
&=\frac{\hbar^2k^2}{2m}+\frac{e^2\rho_1}{2}\left[\left(1+\frac{k}{k_F}\right)\ln\left(1+\frac{k}{k_F}\right)+\left(1-\frac{k}{k_F}\right)\ln\left(1-\frac{k}{k_F}\right)\right]\:,\nonumber
\end{align}
\\
where $k_F=\pi \rho_1/2$ is the largest occupied value of $k$ at $T=0$, such that:

\begin{equation}
\label{Tstraf0}
\mathcal{T}_0(k_F)=\lim_{\beta\rightarrow\infty}\mu=\epsilon_F=\frac{\mathrm{h}^2\rho_1^2}{64m}+\rho_1e^2\ln 2\:.
\end{equation}
\\
For the present aims, the Fermi level $\epsilon_F$ plays the same role as the (false) BEC temperature $T_f$ in the bosonic case. If $\epsilon_F<<\kappa T_c$, the CL approximation can be assumed for $T>T_c$, which yields the axial and radial pressures:

\begin{subequations}
\label{PBzf,PBxf,muf}
\begin{equation}
\label{PBzf}
P_z=-\frac{\kappa T}{S}\frac{\partial}{\partial L}\int_0^\infty\mathrm{d}\epsilon g(\epsilon)\log\left[1+\mathrm{e}^{-\beta(\epsilon-\mu)}\right]\:,
\end{equation}
\\
\begin{equation}
\label{PBxf}
P_x=-\frac{\kappa T}{S}\frac{\partial}{\partial S}\int_0^\infty\mathrm{d}\epsilon g(\epsilon)\log\left[1+\mathrm{e}^{-\beta(\epsilon-\mu)}\right]\:,
\end{equation}
\\
where $\mu$ is determined by the equation:

\begin{equation}
\label{muf}
1=\frac{1}{N}\int_0^\infty\frac{g(\epsilon)}{\mathrm{e}^{\beta(\epsilon-\mu)}+1}\:,
\end{equation}
\end{subequations}
\\
and $g(\epsilon)$ is the semiclassical DOS shown in Eq. \eqref{g(eps)}. Using the numerical solution of Eq. \eqref{muf} in Eq.s \eqref{PBzf} and \eqref{PBxf}, with the same parametric values as in the bosonic case, the results are practically indistinguishable from those reported in Fig.s 3, 4, including the drastic change of the radial pressure at $T_c$. This follows from the semiclassical approximation adopted, in which the non degeneration limit cancels the difference between Fermions and Bosons. However, the parametric values used for Bosons are hardly appliable to Fermions, in concrete cases. Actually, from Eq.s \eqref{Tstraf0} and $u_{eff}=u_0-u_{rep}$, the condition $\epsilon_F<<\kappa T_c=u_{eff}/2$ for the drastic change of the radial pressure to be observable, reads:

\begin{equation}
\nonumber
\frac{\mathrm{h}^2\rho_1^2}{64m}+\rho_1e^2\ln 2<<\frac{u_0-3e^2\rho_1}{2}\,
\end{equation}
\\
which reduces to $e^2\rho_1<<u_0$, if the first term in the l.h.s. is small compared to the second one, i.e., if $\widetilde{\rho}_1<<10^{12}/\widetilde{m}$ (note that $\widetilde{m}\ge1$). This is the case, if $\widetilde{u}_0\lesssim10^3$ (a current condition in vacuum pumps), since:

\begin{equation}
\label{PCF}
e^2\rho_1<<u_0\quad\Rightarrow\quad\widetilde{\rho_1}<<10^{10}\:.
\end{equation}
\\
As in the bosonic case, the condition $\widetilde{u}_{eff}\approx10^{-(1\div2)}$ ensures a confortable range of temperatures ($T_c\lesssim\text{room temperature}$) for observing the partial condensation. Under the condition \eqref{PCF}, however, one has $u_{eff}\approx u_0$, which yields $\widetilde{\rho_1}<<10^{5\div6}$, i.e. a number density $\rho_3=\rho_1/(\pi R^2)$ much smaller than $10^{5\div6}\mathrm{cm}^{-3}$, for a chamber radius $R\approx1\mathrm{cm}$. This pushes the range of observability below the UHV limits. This issue is due to the \emph{exchange} interaction term $u_z(k)$ (Eq. \eqref{uxif2}), which overhelms the free-particle kinetic energy at low temperatures, since the exclusion principle forces the Fermions to occupy the excited states up to $\epsilon_F$. This is not the case for Bosons, of course.

\section{Conclusions}
\label{Concl}

The studies on logarithmic traps (LT's) date back to 1963 \cite{Hoov}, and have been mainly focused on general theoretical aspects \cite{Dechant,Guarnieri,Bouchaud,Aghion1,Aghion2,CEK}, or on possible applications in subnuclear  \cite{Mur1,Mur2,Paa}, or cosmological \cite{Shak} contexts. As for the standard laboratory scales, the most immediate application refers to orbitrons and ultravacuum pumps, which actually produce \emph{electrostatic} logarithmic traps (ELT's) for charged particles. This, however, involves the Coulombic particle-particle interactions, in contrast to the \emph{magnetic} LT's, used to confine neutral particles \cite{PS,Ph}. 

In the present work, we have studied the thermodynamics of a ionic gas, confined in a cylindric chamber (Fig. 1), schematically reproducing the core of an orbitron or ultravacuum pump. Following the Hartree-Fock (HF) procedure at the mean field (MF) level, it is shown that the Coulombic ion-ion interactions produce two main low-temperature effects, on the single-particle energy: a logarithmic repulsive potential $u_x(x)=-u_{rep}\ln(x/x_c)$ (anti-trap), contrasting the ELT, and a kinetic term $u_z(k)$, increasing the free-particle energy $\hbar^2k^2/(2m)$ along the axis itself. The former is a semiclassical consequence of the \textquoteleft coat\textquoteright$\:$ of ions wrapping the axial cathode; the latter is a pure quantum effect, due to the \emph{exchange} interactions.\footnote{To be rigorous, one third of the anti-trap strength too is due to the exchange interactions.}

In standard operative conditions of orbitrons and ultravacuum pumps, the influence of the temperature's changes on the ionic gas can be ignored, which is implicitly accepted without special emphasis or remarks, in the current literature \cite{SD-M,Petit,Cyb}. The present study, instead, shows that such a sort of zero-temperature condition is far from trivial, and follows from what the author called, in (I), the \textquoteleft weirdness\textquoteright$\:$ of the LT's, i.e. their marked tendency to lock the gas (fermionic or bosonic) in the lowest-energy states, at \emph{any} temperature. Actually, it is seen that the temperature's changes are observable only in the ultra vacuum (UHV) regime. Despite a true BEC is excluded, in the UHV regime the radial pressure exerted by a gas of \emph{bosonic} ions on the chamber's circular wall (the anode in Fig. 1), exhibits the behavior shown in Figure 4: a semi-classical slope

\begin{equation}
P_x\approx\rho_3(T-T_c)\:,
\end{equation}
\\
with the absolute zero shifted to the critical temperature $T_c=u_{eff}/(2\kappa)$, and an exponential fall below $T_c$. This is the consequence of the abrupt (but not critical) transition from a non-degenerate to a strongly degenerate regime, occurring when the ions collapse in the radially localized states about the axial cathode. Due to the pure quantum term $u_z(k)$, instead, a gas of \emph{fermionic} ions would show the same effects at pressures well below UHV.

It is interesting to stress that very similar results were obtained in a quite different context of polyelectrolytes generating a LT for water-soluted counterions \cite{Manning}. A collapse of the counterions is predicted to occur just at the same temperature $T_c$, as in the present work, apart from the water dielectric constant, accounting for the chemical environment studied in ref. \cite{Manning} (see Eq. (8) therein). The reason is right the LT's \textquoteleft weirdness\textquoteleft, embrionally mentioned in an informal discussion between the author (G.S. Manning), and L. Onsager (see ref. (13) therein).

\begin{appendices}
\numberwithin{equation}{section}

\section{Calculation of $u_{el}(x,\:z)$}
\label{Appendix A}

Both integrals in $z'$ appearing in Eq.s \eqref{uel0} are in the form:

\begin{equation}
\nonumber
I_a(x,\:z):=\int_{-L/2}^{L/2}\frac{\mathrm{d}z'}{\sqrt{(z-z')^2+X_a^2(x,\:\theta)}}\quad(a=wire,\:wall)\:,
\end{equation}
\\
with

\begin{equation}
\label{A1}
X_a(x,\:\theta):=
\begin{cases}
x\quad&(a=wire)\\
\\
\sqrt{(R-x)^2+4Rx\sin^2(\theta/2)}\quad&(a=wall)\:,
\end{cases}
\end{equation}
\\
and can be easily transformed as follows:

\begin{align}
I_a(x,\:z)&=\int_{-LZ_+}^{LZ_-}\frac{\mathrm{d}\ell}{\sqrt{\ell^2+X_a^2(x,\:z)}}=\nonumber\\
\nonumber\\
&=\frac{1}{2}\ln\left(\frac{\sqrt{1+\Delta_-}+1}{\sqrt{1+\Delta_-}-1}\times\frac{\sqrt{1+\Delta_+}+1}{\sqrt{1+\Delta_+}-1}\right)=\label{A2}\\
\nonumber\\
&=\frac{1}{2}\left[\ln\left(\frac{16}{\Delta_+\Delta_-}\right)+\circ(\Delta_\pm)\right]\:\nonumber,
\end{align}
\\
where

\begin{equation}
\label{A3}
Z_\pm(z):=\frac{1}{2}\pm\frac{z}{L}\quad;\quad\Delta_\pm:=\frac{X_a^2(x,\:z)}{L^2Z_\pm^2}
\end{equation}
\\
For $z\lesssim L/2$, the $\Delta_\pm$'s are small to order $(x/L)^2$ (wire) or to order $Rx/L^2$ (wall) (a constant apart). On retaining the lowest-order terms in Eq. \eqref{A2}, from the definitions \eqref{A3}, it follows that:

\begin{equation}
\label{A4}
I_{wire}(x,\:z)=2\ln\left(\frac{L}{x}\right)+\ln\left(1-\frac{4z^2}{L^2}\right)+\circ(x^2/L^2)\:.
\end{equation}
\\
The calculation of $u_{wall}(x,\:z)$, under the same conditions, involves the integral:

\begin{align}
&J_{wall}(x,\:z):=\frac{1}{2\pi}\int_0^{2\pi}\mathrm{d}\theta I_{wall}=\frac{1}{4\pi}\int_0^{2\pi}\mathrm{d}\theta \left[\ln\left(\frac{16}{\Delta_+\Delta_-}\right)+\circ(Rx/L^2)\right]=\nonumber\\
\label{A5}\\
&=\ln\left(1-\frac{4z^2}{L^2}\right)-\frac{1}{2\pi}\int_0^{2\pi}\mathrm{d}\theta\ln\left(\frac{(R-x)^2+4Rx\sin^2(\theta/2)}{L^2}\right)+\circ(Rx/L^2)\:.\nonumber
\end{align}
\\
It is easy to see that the integral in Eq. \eqref{A5} equals $2\ln(R/L)+\circ(x/R)$, whence, from Eq. \eqref{A4}:

\begin{equation}
u_{el}(x,\:z)=\frac{u_0}{2}\left[I_{wire}(x,\:z)-J_{wall}(x,\:z)\right]\nonumber
\end{equation}
\\
equals expression \eqref{uel1} in the main text.

\section{The mean field (MF) Hamiltonian}
\label{Appendix B}

The integrals appearing in Eq. \eqref{umf} are in the form

\begin{equation}
\label{B1}
I(\mathbf{x},\:z):=\int_{\mathcal{B}}\frac{\mathrm{d}\mathbf{x}'\mathrm{d}z'}{\sqrt{(\mathbf{x}-\mathbf{x}')^2+(z-z')^2}}\phi_{\eta'}(\mathbf{x'})\phi_{\eta}(\mathbf{x}'):
\end{equation}
\\
A series expansion in powers of the components of $\mathbf{x}'$ of initial point $\mathbf{x}'=0$ yields:

\begin{align}
&\frac{1}{\sqrt{(\mathbf{x}-\mathbf{x}')^2+(z-z')^2}}=\frac{1}{\sqrt{x^2+(z-z')^2}}+\frac{\mathbf{x}\centerdot\mathbf{x}'}{[x^2+(z-z')^2]^{3/2}}-\nonumber\\
\nonumber\\
&-\frac{x'^2}{2[x^2+(z-z')^2]^{3/2}}+\frac{3(\mathbf{x}\centerdot\mathbf{x}')^2}{2[x^2+(z-z')^2]^{5/2}}+\circ(x'^3)\:,\nonumber
\end{align}
\\
whence the integral \eqref{B1} reads:

\begin{align}
&I(\mathbf{x},\:z):=\int_{-L/2}^{L/2}\frac{\mathrm{d}z'}{\sqrt{(x^2+(z-z')^2}}\times\Big[\delta_{\eta,\eta'}+\frac{\mathbf{x}\centerdot\langle\:\eta\:|\mathbf{x'}|\:\eta'\:\rangle}{x^2+(z-z')^2}-\nonumber\\
\label{B2}\\
&-\frac{\langle\:\eta\:|x'^2|\:\eta'\:\rangle}{2[x^2+(z-z')^2]}+\frac{3\langle\:\eta\:|(\mathbf{x}\centerdot\mathbf{x'})^2|\:\eta'\:\rangle}{2[x^2+(z-z')^2]^2}+\circ(\langle\:\eta\:|x'^3|\:\eta'\:\rangle)\Big]\:,\nonumber
\end{align}
\\
where 

\begin{equation}
\label{B3}
\langle\:\eta\:|\cdots|\:\eta'\:\rangle=\int\mathrm{d}\mathbf{x}'\phi_\eta(\mathbf{x}')\cdots\phi_{\eta'}(\mathbf{x}')
\end{equation}
\\
are real matrix elements. Note that the second term in square brakets vanishes if $\phi_\eta(\mathbf{x}')$ and $\phi_{\eta'}(\mathbf{x}')$ have the same parity (in particular if $\eta=\eta'$), while the third and fourth terms vanish if the parity is opposite. The substitution $\theta=(z-z')/x$ in the three corrective terms leads to the expression:

\begin{align}
&I(\mathbf{x},\:z)=\overbrace{\int_{-L/2}^{L/2}\frac{\mathrm{d}z'}{\sqrt{(x^2+(z-z')^2}}\delta_{\eta,\eta'}}^{I_0(\mathbf{x},\:z)}+\label{B4}\\
\nonumber\\
&+\frac{\mathbf{x}\centerdot\langle\:\eta\:|\mathbf{x'}|\:\eta'\:\rangle}{x^2}\int_{-\theta_+}^{\theta_-}\frac{\mathrm{d}\theta}{(1+\theta^2)^{3/2}}-\frac{\langle\:\eta\:|x'^2|\:\eta'\:\rangle}{2x^2}\int_{-\theta_+}^{\theta_-}\frac{\mathrm{d}\theta}{(1+\theta^2)^{3/2}}+\nonumber\\
\label{B5}\\
&+\frac{3\langle\:\eta\:|(\mathbf{x}\centerdot\mathbf{x'})^2|\:\eta'\:\rangle}{2x^4}\int_{-\theta_+}^{\theta_-}\frac{\mathrm{d}\theta}{(1+\theta^2)^{5/2}}+\circ(\langle\:\eta\:|x'^3|\:\eta'\:\rangle)\:,\nonumber
\end{align}
\\
where $\theta_\pm(x,\;z):=(L/2\pm z)/x$ are divergingly large for $z\lesssim L/2$, since $x<R<<L$. In the limit $\theta_\pm\rightarrow\infty$, the correction Eq. \eqref{B5} to $I_0$ (Eq. \eqref{B4}) reads simply:

\begin{align}
&I(\mathbf{x},\:z)-I_0(\mathbf{x},\:z)=\nonumber\\
\label{B6}\\
& =\frac{2\mathbf{x}\centerdot\langle\:\eta\:|\mathbf{x'}|\:\eta'\:\rangle}{x^2}-\frac{\langle\:\eta\:|x'^2|\:\eta'\:\rangle}{x^2}+\frac{2\langle\:\eta\:|(\mathbf{x}\centerdot\mathbf{x'})^2|\:\eta'\:\rangle}{x^4}+\circ(\langle\:\eta\:|x'^3|\:\eta'\:\rangle)\:.\nonumber
\end{align}
\\
In order to estimate the matrix elements \eqref{B3}, let $f_\eta(\mathbf{y})$ be a dimensionless rescaled expression of $\phi_\eta(\mathbf{x})$, in the dimensionless variable $\mathbf{y}=\mathbf{x}/\ell_\eta$ (D=2):

\begin{equation}
\nonumber
f_\eta(\mathbf{x}/\ell_\eta):=\ell_\eta\phi_\eta(\mathbf{x})\quad\Rightarrow\quad\int\mathrm{d}\mathbf{y}f_\eta^2(\mathbf{y})=1\:.
\end{equation}
\\
Since $\ell_\eta$ is the localization length of the radial state $\phi_\eta(\mathbf{x})$, the condition $y\lesssim1$ determines the range where $f_\eta$ does efficiently contribute any integral in $\mathbf{y}$. For the present aims, we can limit ourselves to the study of

\begin{align}
&\langle\:\eta\:|x'^\alpha|\:\eta'\:\rangle=\nonumber\\
\nonumber\\
&=\int\mathrm{d}\mathbf{x}'x'^\alpha\phi_\eta(\mathbf{x}')\phi_{\eta'}(\mathbf{x}')=\frac{(\ell_\eta)^{\alpha+1}}{\ell_{\eta'}}\int\mathrm{d}\mathbf{y}y^\alpha f_\eta(\mathbf{y})f_{\eta'}(\mathbf{y\ell_\eta/\ell_{\eta'}})\:,\label{B7}
\end{align}
\\
with the substitution $\mathbf{x}\rightarrow\mathbf{y}\ell_\eta$ and $\alpha>0$. If $\ell_\eta\approx\ell_{\eta'}$, the integral in Eq. \eqref{B7} is a numerical factor of order unity, depending on $\alpha$, $\eta$, $\eta'$, which yields $\langle\:\eta\:|x'^\alpha|\:\eta'\:\rangle\approx(\ell_\eta)^\alpha\approx(\ell_{\eta'})^\alpha$. If, instead, $\ell_\eta>>\ell_{\eta'}$, the function $f_{\eta'}(\mathbf{y}\ell_\eta/\ell_{\eta'})$ contributes the integral only for $y\lesssim\ell_{\eta'}/\ell_\eta$. Hence the integral itself turns out to be of order $(\ell_{\eta'}/\ell_{\eta})^{\alpha+1}$. As far as the order of magnitude is concerned, what precedes yields:

\begin{equation}
\label{B8}
\langle\:\eta\:|x'^\alpha|\:\eta'\:\rangle\propto\left(\min\{\ell_{\eta'},\:\ell_{\eta}\}\right)^\alpha\:.
\end{equation}
\\
When applied to Eq. \eqref{B6}, equation \eqref{B8} leads to the statement of the main text, concerning the order of the approximation $\mathbf{x}'=0$ in Eq. \eqref{umf}.

\section{The semi-classical DOS}
\label{Appendix C}

For brevity, in the present appendix the symbol $u_{eff}$ will be replaced by $u$. 

Given the DOS's $g_z(\epsilon')$ (Eq. \eqref{gz2}) and $g_x(\epsilon'')$ (Eq. \eqref{gx2}), the total DOS for the energy $\epsilon=\epsilon'+\epsilon''$ reads:

\begin{subequations}
\begin{align}
g(\epsilon)&=\int_0^\epsilon\mathrm{d}\epsilon'g_z(\epsilon')g_x(\epsilon-\epsilon')=\frac{L\sqrt{2m}}{\mathrm{h}}\int_0^\epsilon\frac{\mathrm{d}\epsilon'}{\sqrt{\epsilon'}}g_x(\epsilon-\epsilon')\label{C1a}\\
\nonumber\\
&=\frac{L\sqrt{2m}\widehat{S}}{4\mathrm{h}\sqrt{u}}\mathrm{e}^{2(\epsilon-\epsilon_c+\alpha_0)/u}\underbrace{\int_0^\epsilon\frac{\mathrm{d}\epsilon'}{\sqrt{\epsilon'}}\mathrm{e}^{-2\epsilon'/u}}_{\mathrm{Erf(\sqrt{2\epsilon/u})}\sqrt{u\pi/2}}\quad(\epsilon\le\epsilon_c)\:,\label{C1b}
\end{align}
\end{subequations}
\\
since $\epsilon\le\epsilon_c$ necessarily impies $\epsilon-\epsilon'\le\epsilon_c$. For $\epsilon\ge\epsilon_c$ equation \ref{C1a} involves two separate integrals, corresponding to the two cases $\epsilon-\epsilon'\lessgtr\epsilon_c$:

\begin{align}
&g(\epsilon)=\frac{L\sqrt{2m}\widehat{S}}{4\mathrm{h}\sqrt{u}}\mathrm{e}^{2\alpha_0/u}\left[\int_0^{\epsilon-\epsilon_c}\frac{\mathrm{d}\epsilon'}{\sqrt{\epsilon'}}g_x(\epsilon-\epsilon')+\int_{\epsilon-\epsilon_c}^\epsilon\frac{\mathrm{d}\epsilon'}{\sqrt{\epsilon'}}g_x(\epsilon-\epsilon')\right]=\nonumber\\
\label{C2}\\
&=\frac{L\sqrt{2m}\widehat{S}}{4\mathrm{h}u}\mathrm{e}^{2\alpha_0/u}\left[2\sqrt{\epsilon-\epsilon_c}+\mathrm{e}^{2(\epsilon-\epsilon_c)/u}\int_{\epsilon-\epsilon_c}^\epsilon\frac{\mathrm{d}\epsilon'}{\sqrt{\epsilon'}}\mathrm{e}^{-2\epsilon'/u}\right]\:.\nonumber
\end{align}
\\
By representing the second integral in Eq. \ref{C2} as in Eq. \ref{C1b} (see underbrace), one easily gets:
 
\begin{align}
&g(\epsilon)=\overbrace{\frac{L\sqrt{2m}}{\sqrt{2}\mathrm{h}u}\mathrm{e}^{2\alpha_0/u}}^{\mathcal{G}}\times\Big[\widehat{S}\sqrt{\epsilon-\epsilon_c}\:+\nonumber\\
\label{C3}\\
&+\sqrt{\frac{u\pi}{8}}\mathrm{e}^{2\epsilon/u}\left(\mathrm{Erf}(\sqrt{2\epsilon/u})-\mathrm{Erf}(\sqrt{2(\epsilon-\epsilon_c)/u})\right)\Big]\quad(\epsilon\ge\epsilon_c)\:.\nonumber
\end{align}
\\
Equations \eqref{C1b} and \eqref{C3} correspond to Eq.s \eqref{g(eps)} in the main text.

\end{appendices}

\medskip

The author reports there are no competing interests to declare

\end{document}